\documentstyle[epsfig,cite]{article}    % Specifies the document style.
 
\large
\newcommand{\Title}[1]{\begin{center}{\Large\bf #1}
\end{center}\vskip 0.1in}
\newcommand{\Name}[1] {\begin{center}{\large    #1}
\end{center}\vskip 0.1in}

\def\Thebibliography#1{\section*{REFERENCES}\list
 {[\arabic{enumi}]}{\settowidth\labelwidth{[#1]}\leftmargin\labelwidth
 \advance\leftmargin\labelsep
 \usecounter{enumi}}
 \def\newblock{\hskip .11em plus .33em minus .07em}
 \sloppy\clubpenalty4000\widowpenalty4000
 \sfcode`\.=1000\relax}

\newcommand{\FigureAng}[4]{
  \mbox{\epsfig{file=#1,width=#2,height=#3,angle=#4}}
  }
\def\Journal#1&#2&#3(#4){\unskip, #1~{\bf #2} (#4) #3}
\def\NPB{Nucl.\ Phys.\ B}

\def\PLB{Phys.\ Lett.\ B}
\def\PRL{Phys.\ Rev.\ Lett.}
\def\PRD{Phys.\ Rev.\ D}

\def\ZPC{Z.\ Phys.\ C}

\def\PTP{Prog. \ Theor. \ Phys.}
\def\SJNP{Sov. \ J. \ Nucl. \ Phys.}
\def\etal{{\it et al.}}

\def\b0b0{${\rm B^{o}\overline{B^o}}$}

\newcommand{\be}{\begin{equation}}
\newcommand{\ee}{\end{equation}}
\newcommand{\ba}{\begin{array}{c}}
\newcommand{\ea}{\end{array}}
\newcommand{\beqn}{\begin{eqnarray}}
\newcommand{\eeqn}{\end{eqnarray}}

\def\beq{\begin{equation}}
\def\eeq{\end{equation}}
\newcommand{\bea}{\begin{eqnarray}}
\newcommand{\eea}{\end{eqnarray}}
\newcommand{\nn}{\nonumber}
\def\simleq{\; \raise0.3ex\hbox{$<$\kern-0.75em
      \raise-1.1ex\hbox{$\sim$}}\; }
\def\simgeq{\; \raise0.3ex\hbox{$>$\kern-0.75em
      \raise-1.1ex\hbox{$\sim$}}\; }
\def\noi{\noindent}

\def\R{ {\rm R \kern -.31cm I \kern .15cm}}
\def\C{ {\rm C \kern -.15cm \vrule width.5pt
\kern .12cm}}
\def\Z{ {\rm Z \kern -.27cm \angle \kern .02cm}}
\def\N{ {\rm N \kern -.26cm \vrule width.4pt \kern .10cm}}
\def\1{{\rm 1\mskip-4.5mu l} }
 
\begin{document}

\large
%\tableofcontents
% \listoftables
% \listoffigures
 
\Title{Measuring $|V_{ub}|$ with
B$\rightarrow D_s^+ X_u$ transitions }
 
\Name{{\bf R. Aleksan, M. Zito} \\
Commissariat \`a l'Energie Atomique, Saclay, \\
DSM/DAPNIA/SPP \\
91191 Gif-sur-Yvette Cedex, France}  

\Name{{\bf A. Le Yaouanc, L. Oliver, O. P\`ene and J.-C. Raynal} \\
Laboratoire de Physique Th\'eorique\footnote{Unit\'e Mixte de Recherche CNRS - UMR
8627} \\ Universit\'e de Paris XI, B\^at. 211\\
91405 Orsay Cedex, France }
\vskip 1 truecm
\centerline{\bf Abstract} \par \vskip 3 truemm

We propose the determination of the CKM matrix element  $|V_{ub}|$ by the measurement
of the spectrum of $B \to D_s^+ X_u$, dominated by the spectator quark model
mechanism $\bar{b} \to D_s^{(*)+} \bar{u}$. The interest of considering $B \to D_s^+X_u$ versus
the semileptonic decay is that more than 50 \% of the spectrum for $B \to D_s^+ 
X_u$ occurs above the kinematical limit for $B \to D_s^+  X_c$, while most of the
spectrum $B \to l \nu  X_u$ occurs below the $B \to l \nu  X_c$ one. Furthermore, the
measure of the hadronic mass $M_X$ is easier in the presence of an identified $D_s$ than
when a $\nu$ has been produced. As a consistency check, we point out that the rate
$\bar{b} \to D_s^{(*)+}  \bar{c}$ (including QCD corrections that we present elsewhere) is
consistent with the measured $BR (B \to D_s^{\pm}  X)$. Although the hadronic
complications may be more severe in the mode that we propose than in the semileptonic
inclusive decay, the end of the spectrum in $B \to l \nu  X_u$ is
not well understood on theoretical grounds. We argue that, in our case,
the excited $D_s^{**}$, decaying into $D  K$, do not contribute and, if there is tagging of
the $B$ meson, the other mechanisms to produce a $D_s$ of the right sign are presumably 
small, of $O(10^{-2})$ relative to the spectator amplitude, or can be controlled by
kinematical cuts.   In the absence
of tagging, other hadronic backgrounds deserve careful study. We present a feasability study with the BaBar detector. \par  

\vskip 1 truecm \noindent LPT-Orsay 99-35 \par 
\noindent DAPNIA/SPP 99-18 \par
\noindent  May 1999

\newpage
\pagestyle{plain}
\baselineskip=26 pt 
\section{Introduction}
\hspace{\parindent} 
The determination of the strength of the transition between b and u quarks
is a very important goal for understanding the sector of the theory 
involving flavor mixing.
Indeed, the value of the element $|V_{ub}|$ in the Cabibbo-Kobayashi-Maskawa
(CKM) mixing matrix \cite{CKM} is a key ingredient which is used
to determine the unitarity triangle and thus test the consistency 
of the Standard Model in the sector responsible for CP violation.
It is also one of the most difficult measurements in B physics, in particular
due to the large and model dependent theoretical uncertainties.
The methods which have been used so far to extract $|V_{ub}|$ involve 
semileptonic B decays.
The first method uses
the inclusive lepton spectrum above the kinematical limit for
$b \rightarrow c$ transitions while the second technique requires
the exclusive reconstruction of $B\rightarrow \pi l\nu$ or $\rho l\nu$. 
The errors in the first case are due to the fact that only a tiny fraction
of the lepton energy spectrum from $b \rightarrow u l\nu$ is observed, that parton
model evaluation is questionable in this kinematical region and that a large model
dependent extrapolation is necessary to extract the total rate. An improvement based on
studying the hadronic mass spectrum increases the signal but is not free of problems
related to the $b \to c$ background \cite{1bisr}. In the
second case, the uncertainties are mainly due to the limited  statistics and
the theoretical uncertainty in the form factors for the $B\rightarrow \pi$ and 
$B\rightarrow \rho$ transitions.

We would like in the following to propose a new approach to measure $|V_{ub}|$ which 
involves inclusive $ B\rightarrow D^+_s$ transitions where
we make use as much as possible of experimentally measured 
parameters in order to reduce the uncertainties.
In these decays the $D_s$ meson is essentially produced via the virtual W 
emitted 
by the b quark (see figure~\ref{fig:diagsp}). 
We shall discuss later the other possibilities to produce 
a $D_s$ meson and make a preliminary survey of the backgrounds and hadronic
uncertainties  of our method to measure $|V_{ub}|$. The $b \rightarrow u$ transitions
are identified by requiring the momentum of the $D_s$ meson to be in the range
above the kinematical  limit for the decay 
$ B\rightarrow D^+_s \overline{D}$ (i.e. $\sim 1.82$ GeV 
in the B meson center of mass) and up to 2.27 GeV corresponding to the 
transition $B\rightarrow D^+_s \pi$. 
It is very important to note here that in contrast to the 
inclusive semileptonic case {\it this range includes 
the majority of the $\overline{b}\rightarrow D^+_s \bar{u}$ transitions}
and therefore a smaller extrapolation is needed to obtain the
total rate. Of course, a drawback of this new method is that,
since it concerns purely hadronic transitions, it is subject to other hadronic
uncertainties than the semileptonic end spectrum $B \to l \nu X_u$. After calculating
the inclusive rate for  $ B\rightarrow D^+_s X_q$ we discuss how $|V_{ub}|$ is
extracted and then enumerate and try to estimate the uncertainties in section 3.
Various sources of  background are studied and  rejection methods are
proposed in section 4 for tagged events and in section 5 for untagged events.
Finally, in section 6 we present a feasability study for the BaBar detector, and in
section 7 we conclude.\par

 When this paper was finished, we noticed that other methods to
measure $V_{ub}$ have been proposed using channels that involve also the
$(\bar{s}c)(\bar{u}b)$ weak coupling. Namely, the totally {\it inclusive} $B$ decays
through $b \to \bar{c}su$ has been proposed \cite{2r} or rare exclusive decays of the
type $B^+ \to D_s^+\gamma$ \cite{3r}. However, although the weak coupling is the same,
these methods do not overlap with the proposition of our paper to measure $|V_{ub}|$.

\section{The $B \rightarrow D^+_s X_q$ rate} \hspace{\parindent}
The inclusive decay rate of a B meson decaying into a $D^+_s$ meson 
is obtained using the spectator quark model by writing

\be
\Gamma (B \rightarrow D^+_s X_q) \simeq 
\Gamma (\overline{b}\rightarrow D^+_s \overline{q}) + 
\Gamma (\overline{b}\rightarrow D^{*+}_s \overline{q})
\ee
where $q$ is the outgoing quark as shown in figure~\ref{fig:diagsp} 
(other diagrams exists
and will be discussed later). One should note that decays to the lowest P wave
$D^{**}_s$ states do not lead to $D_s$ mesons since their main decays
are $D^{**}_s \rightarrow D^{(*)} K $.
%%%%%%%%%%%%%%%%%%% Figure %%%%%%%%%%%%%%%%%%%%%%%%%%
\setlength{\unitlength}{0.7mm}
\begin{figure*}[htb]
\vfill
\begin{picture}(280,80)(-50,-80)
%\begin{center}
\FigureAng{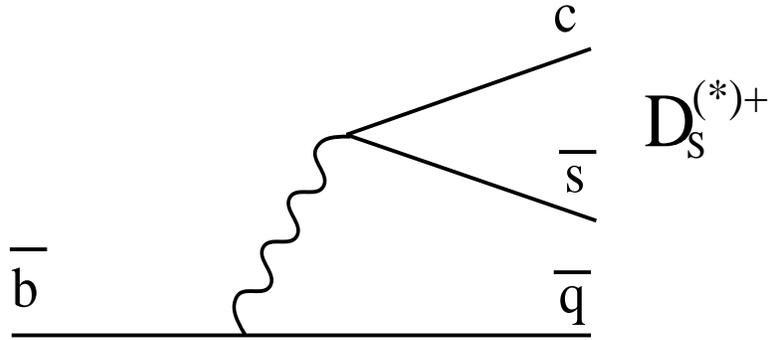}{5.0cm}{10.0cm}{-90}
%\FigureAng{btods_fig1.eps}{9.0cm}{12.0cm}{90}
%\mbox{\epsfig{file=/home/usr201/mnt/aleksan/btods_fig1.eps}}
%\end{center}
\end{picture}
\vfill
\caption{Spectator diagram for the decay 
$\overline{b}\rightarrow D^+_s \overline{q}$. }
\label{fig:diagsp}
\end{figure*}
%%%%%%%%%%%%%%%%%%%% Figure Unitarity Triangle %%%%%%%%%%%%%

Extending the standard vacuum insertion approximation, successful in exclusive
decays, the effective matrix element used for the weak decay  $\overline{b}\rightarrow
D^{(*)+}_s \overline{q}$ reads \be
\langle D^{(*)+}_s \overline{q} | {\cal H}_{eff} | \overline{b} \rangle \ =
\ \frac{G_F}{\sqrt{2}} a_1 V^*_{qb}V_{cs}
\langle D^{(*)+}_s | A^\mu(V^\mu) | 0\rangle 
\langle \overline{q} | J_{\mu q} | \overline{b} \rangle
\label{eq:hamil}
\ee
where $G_F$ is the Fermi constant, $V_{ij}$ are the Cabibbo-Kobayashi-Maskawa
matrix elements and
\be
a_1 = c_1+\frac{c_2}{N_c}
\label{eq:jq}
\ee
\noindent  is a combination of short distance QCD factors, and the current $J_{\mu q}$
reads~:
 \be
J_{\mu q} = \overline{q}\gamma_\mu (1-\gamma_5) b \quad . 
\ee
\noindent  We have, for the emission of a pseudoscalar~: 
\be
\langle D^+_s | A^\mu | 0\rangle = -i f_P p^\mu_P\ \  
\ee
\noindent  and for the emission of a vector meson
\be
\langle D^{*+}_s | V^\mu | 0\rangle = m_V f_V \epsilon^{*\mu}_V\ \ 
 \ee 

Here $\epsilon^*$ is the polarisation 
quadrivector of the meson. In Eq.~\ref{eq:jq}, the Wilson coefficients
are \cite{GLAM} 
\be
\label{7e}
c_1 = \frac{c_+ +c_-}{2}\ \ \mbox{and}\ \  c_2 = \frac{c_+ -c_-}{2}\ ,\ \ 
c_\pm = \left[\frac{\alpha_s(\mu )}{\alpha_s(m_W )}\right]^{d_\pm}
\ee
where $d_+=-6/23$ and $d_-=12/23$. In writing Eq.~\ref{eq:hamil}, 
factorization has been assumed. This assumption is justified since
the diagram involved here (Fig.~\ref{fig:diagsp}) is the spectator diagram
with external emission of the W. Indeed no internal emission diagram 
nor penguin diagrams exist. Factorization is so far
consistent with the experimental data in exclusive decays where only Fig.~\ref{fig:diagsp} type
diagrams are involved and the parameter $|a_1|= 1.00 \pm 0.06$ 
has been extracted using a combined fit of se\-ve\-ral measured modes. 
On the other hand, factorization, up to calculable corrections, has been
proved recently in the $m_b \to \infty$ limit for $B \to \pi \pi$ \cite{6r}. However
one should be aware that it has been shown \cite{SV} that duality between the parton
model and the sum over all exclusive  channels with the factorization assumption
may in general have corrections at the 1/$N_c$ level. Discussions on this point can
be found in refs.~\cite{ALOPR}.

The width of the inclusive $\overline{b}\rightarrow D^+_s \overline{q}$ 
is calculated easily 
at the tree level by evaluating the diagram in figure~\ref{fig:diagsp}.
One finds:
\be
\Gamma^{(0)} (\overline{b}\rightarrow D^+_s \overline{q}) =
\frac{G_F^2}{8\pi} |V^*_{qb}V_{cs}|^2 f_{D_s}^2 
\frac{(m_b^2-m_q^2)^2}{m_b^2} \left( 1 - 
\frac{m_{D_s}^2(m_b^2+m_q^2)}{(m_b^2-m_q^2)^2} \right)
p_{D_s}a_1^2
\label{eq:width}
\ee
where $p_{D_s} = \sqrt{[m_b^2-(m_{D_s}+m_q)^2][m_b^2-(m_{D_s}-m_q)^2]}/2m_b$ 
is the momentum of the outgoing $D_s$ meson in the b rest frame,
$G_F$ is the Fermi constant and $f_{D_s}$ is the $D_s$ decay constant.
The notation $\Gamma^{(0)}$ is used for the width without including the
radiative corrections.
A similar formula is obtained for 
$\Gamma^{(0)}(\overline{b}\rightarrow D^{*+}_s (\lambda =0)\ \overline{q})$
where the $D^{*+}_s$ is longitudinally polarized
by replacing in Eq.~\ref{eq:width}.
$f_{D_s}$ with $f_{D^*_s}$, $m_{D_s}$ with $m_{D^*_s}$.
For the transverse polarization ($\lambda =\pm 1$) we find
\be
\Gamma^{(0)} (\overline{b}\rightarrow D^{*+}_s(\lambda =\pm 1)\ \overline{q}) =
\frac{G_F^2}{4\pi} |V^*_{qb}V_{cs}|^2 f_{D^*_s}^2 m_{D^*_s}^2 
\frac{m_b^2+m_q^2}{m_b^2} \left( 1 - 
\frac{m_{D^*_s}^2}{m_b^2+m_q^2} \right)
p_{D^*_s}a_1^2
\ee
It is interesting to note that neglecting $m_{D^*_s}^2$ compared to $m_b^2$
and for $m_q^2 << m_b^2$, one obtains
\be
\frac{\Gamma_T}{\Gamma_L} =
\frac{\Gamma^{(0)}(\overline{b}\rightarrow D^{*+}_s(\lambda =\pm 1)\ \overline{q})}
{\Gamma^{(0)}(\overline{b}\rightarrow D^{*+}_s (\lambda =0)\ \overline{q})} 
\simeq \frac{2m_{D^*_s}^2}{m_b^2-4m_q^2}
\ee
and therefore transverse polarizations are suppressed.
\begin{table}[hbt]
\begin{center}
\begin{tabular}{|c|c|c|}
\hline
 & $\overline{b}\rightarrow D^{*+}_s \overline{c}$ &
$\overline{b}\rightarrow D^{*+}_s \overline{u}$ \\ \hline
$\Gamma_T/\Gamma_L$ & $\sim 1/2$ & $\sim 1/3$ \\ \hline
\end{tabular}
\end{center}
\caption{ Fraction of tranverse to longitudinal polarized $D_s^*$ mesons in 
inclusive decays.}
\label{tab:polar}
\end{table}
As an illustration, Table~\ref{tab:polar} shows the expected order of magnitude of the ratio $\Gamma_T/\Gamma_L$
for $\overline{b}\rightarrow D^{*+}_s \overline{c}$ and
$\overline{b}\rightarrow D^{*+}_s \overline{u}$ transitions.  Experimental verifications of
table~\ref{tab:polar}  would be useful and would give
further confidence in the method proposed here.
Adding both longitudinal and transverse polarizations, one has
\be
\Gamma^{(0)} (\overline{b}\rightarrow D^{*+}_s \overline{q}) =
\frac{G_F^2}{8\pi} |V^*_{qb}V_{cs}|^2 f_{D^*_s}^2 
\frac{(m_b^2-m_q^2)^2}{m_b^2} \left( 1 + 
\frac{m_{D^*_s}^2(m_b^2+m_q^2-2m_{D^*_s}^2)}{(m_b^2-m_q^2)^2} \right)
p_{D^*_s}a_1^2
\ee

>From (\ref{eq:jq}) and (\ref{7e}) it can be seen that the {\it short distance} QCD
factor $a_1 = 1 + O(\alpha^2_s)$, i.e. the correction to the tree rate is of second
order in $\alpha_s$. We have computed elsewhere \cite{7r} the radiative corrections
to $b \to D_s^{(*)-}u$ at order $\alpha_s$, that involve vertex, self-energy and
Bremsstrahlung diagrams. These radiative corrections are evaluated at the order 
$\alpha_s$ in the same way than 
for the semileptonic decays \cite{CM,NIR}, i.e. on the lower quark legs in Fig. 1. 
This is because the $D^{(*)}_s$ is a color singlet. We have obtained,
within the on-shell renormalization scheme \cite{7r}~:

\be
\Gamma (\overline{b}\rightarrow D^{(*)+}_s \overline{q}) =
\Gamma^{(0)} (\overline{b}\rightarrow D^{(*)+}_s \overline{q}) \left [ 1 +
{4 \over 3} \ {\alpha_s \over \pi} \ \eta^{(*)} \left ( \xi_{D_s^{(*)}}, r_q
\right ) \right ] \ee 

\noindent where $\xi = {q^2 \over m_b^2}$ and $r_q = {m_q \over m_b}$, with $q^2 =
m_{D_s}^2$ or $m^2_{D_s^*}$. As shown in \cite{7r}, in the limit $\xi \to 0$, $r_q \to
0$ one finds

\be
\eta(0,0) = \eta^*(0,0) = {5 \over 4} - {\pi^2 \over 3}  \quad .\ee

\noindent The functions $\eta^{(*)}(\xi , r)$ are slowly varying with $r$ and $\xi$.
\par  

In order to derive the expected branching fractions,
the following numerical values are used for the pole quark masses, extracted from
an analysis of semi-leptonic $B$ decays at first order, to be coherent with the
present first order calculation (see ref. \cite{7r} for a discussion on the choice of
these parameters)~:   \be m_b = 4.85\ GeV/c^2\ , \ m_c = 1.45\ GeV/c^2 
\ee
and we take $m_u \cong 0$ and the decay constants 
\be 
\ f_{D_s} = 230\ MeV\ \ \ , \ \ \ f_{D^*_s}=280\ MeV
\ee
With these values, and $\alpha_s (m_b) = 0.2$, the radiative
corrections take the following values (the mass dependence is discussed in
\cite{7r}) for $q = c$~:

\bea
&&{4 \over 3} \ {\alpha_s \over \pi} \ \eta\left ( \xi_{D_s}, r_c \right ) = - 0.095
\nn \\ &&{4 \over 3} \ {\alpha_s \over \pi} \ \eta^*\left ( \xi_{D_s^*}, r_c \right )
= - 0.108 \eea

\noi and for $q = u$~:
\bea
&&{4 \over 3} \ {\alpha_s \over \pi} \ \eta\left ( \xi_{D_s}, 0 \right ) = - 0.168 \nn
\\ &&{4 \over 3} \ {\alpha_s \over \pi} \ \eta^*\left ( \xi_{D_s^*}, 0 \right ) = -
0.159 \quad . \eea

Using $\tau_B$ = 1.6 ps and $|V_{cb}|=0.04$, one calculates, including the QCD
corrections

 \be
Br(\overline{b}\rightarrow D^{(*)+}_s \overline{c}) \simeq 8.0\%
\ee
where $Br(\overline{b}\rightarrow D^+_s \overline{c}) \simeq 2.6\%$ and
$Br(\overline{b}\rightarrow D^{*+}_s \overline{c}) \simeq 5.4\%$
and with $|V_{ub}|/|V_{cb}|=0.08$
\be
Br(\overline{b}\rightarrow D^{(*)+}_s \overline{u}) \simeq 6.8\times 10^{-4}
\label{eq:BtoDsu}
\ee
where $Br(\overline{b}\rightarrow D^+_s \overline{u}) \simeq 2.3\times 10^{-4}$
and 
$Br(\overline{b}\rightarrow D^{*+}_s \overline{u}) \simeq 4.5\times 10^{-4}$.

At this stage, several points should be underlined:
\begin{itemize}
\item The sensitivity of the rate to the b quark mass goes as $m_b^3$ 
instead of $m_b^5$ in the case of the semileptonic decay.
\item The sensitivity of the decay rate with respect to the mass $m_q$ 
is negligible for the light quarks. 
It is not dramatic for the c quarks, in particular if $m_b-m_c$ is known to a good
accuracy  (Eq.~\ref{eq:width}). 
\item The calculated overall branching fraction for 
$Br(\overline{b}\rightarrow D^{(*)+}_s \overline{c}) \simeq 8.0\%$ 
is in agreement within 1$\sigma$ with
the value measured\cite{cleo:ds}~: 

\be
BR(B \to D_s^{\pm}X) = (10.0 \pm 2.5)\% \quad .
\ee 
The observed agreement is encouraging as it shows that the very simple
approach at the quark level accounts rather well for the data. 
Equivalently, one could extract $|V_{cb}|$. 
Using $m_b = (5.0\pm 0.20)\ GeV/c^2$ and the relative error 
$\sigma (f_{D^{(*)}_s})/f_{D^{(*)}_s} =0.1$,
we find $|V_{cb}|=0.044\pm 0.008$.
\item On the theoretical level it would be necessary to investigate these inclusive
processes using the $1/m_b$ expansion. On the one hand, one would need to
estimate the next-to-leading non-perturbative corrections. On the other hand, a
systematic analysis of these inclusive hadronic processes $B \to D_s^+X$ could
hopefully give independent information on those non-perturbative parameters of
the heavy quark expansion such as $\overline{\Lambda}$, $\lambda_1$ and
$\lambda_2$ \cite{BSU}.    \end{itemize}

\section{Measurement of $|V_{ub}|$ using $B \overline{B}$ pairs from 
$\Upsilon (4S)$ decays}
\hspace{\parindent}
In a similar way than for the measurement of $|V_{cb}|$, it should be possible 
to determine $|V_{ub}|$ by selecting $D_s$ mesons with momentum above 
the kinematical limit for $ B\rightarrow D^+_s \overline{D}$. 
The $D_s$ momentum in the latter case is 1.82 GeV/c in the B rest frame.
However, for B pair production at the $\Upsilon (4S) $, B mesons are 
generated with a momentum of about 300 MeV/c and 
therefore the latter limit 
is of the order of 2.0 GeV/c as can be seen in figure~\ref{fig:ds_plot}. To extract
$|V_{ub}|$, it is thus necessary to estimate the fraction of 
$B \rightarrow D^+_s X_u$ decays with $p_{D_s} > 2.0$ GeV/c. 
We have computed the expected 
momentum spectrum of $D^+_s$ produced via the spectator diagram in 
figure~\ref{fig:diagsp}, taking into account the $b$-quark Fermi motion inside the
$B$ meson using the ACCMM model \cite{ACCMM} at tree level, neglecting for the moment
the radiative corrections. This spectrum is shown in figure~\ref{fig:vubdsmom}. The striking feature of this distribution is that the average $D_s$ momentum
is above 2.0 GeV/c with about 75\% of the $D_s$ mesons above that limit.
Obviously this fraction depends on the theoretical parameter and 
therefore we have varied $p_F$ in the reasonable range 
($200\ MeV/c < p_F < 400\ MeV/c$) 
to evaluate the possible systematic uncertainties related to that parameter.
Table~\ref{tab:eff_ds} shows the sensitivity of the fraction of $D_s$ for 
various cuts on $p_{D_s}$, assuming different values for $p_F$ and the mass
of the spectator quark.
Should it be possible to measure the recoiling mass to the $D_s^{(*)}$, the
value of the cut on $p_{D_s}$ could be reduced, thus increasing the efficiency.
  
%%%%%%%%%%%%%%%%%%% Figure %%%%%%%%%%%%%%%%%%%%%%%%%%
\setlength{\unitlength}{0.7mm}
\begin{figure*}[htb]
\vfill
\begin{picture}(280,90)(-25,-0)
%\begin{center}
%\FigureAng{vcbdsmom.eps}{5.0cm}{10.0cm}{-90}
%\FigureAng{babardwg.eps}{9.0cm}{12.0cm}{90}
\mbox{\epsfig{file=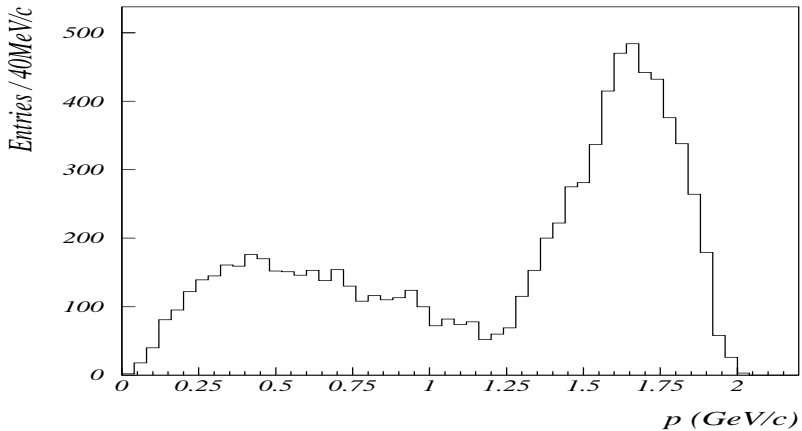,height=7.0cm,width=12cm}}
%\end{center}
\end{picture}
\vfill
\caption{Momentum spectrum for $D^+_s$ mesons produced from the reaction
$\overline{b}\rightarrow D^+_s \overline{c}$ (i.e. upper vertex). Decays 
with a $D^{**}$ meson from the lower vertex have not been included 
in this plot.
These decays tend to fill the slight deep at 1.25 GeV/c but do not affect
the end of the spectrum.}
\label{fig:ds_plot}
\end{figure*}
%%%%%%%%%%%%%%%%%%%% Figure Unitarity Triangle %%%%%%%%%%%%%
\begin{table}[hbt]
\begin{center}
\begin{tabular}{|c|c|c|c|c|c|c|c|c|c|c|}
\hline
$m_u$ $m_d$(MeV) & $p_F$ (MeV) & 
\multicolumn{3}{c|}{$ p_{D_s} > 2 GeV/c $} &
\multicolumn{3}{c|}{$ p_{D_s} > 2.05 GeV/c $} & 
\multicolumn{3}{c|}{$ p_{D_s} > 2.1 GeV/c $}  \\ \hline
 & & $D_s$ & $D_s^*$ & all & $D_s$ & $D_s^*$ & all & $D_s$ & $D_s^*$ & all \\
150 & 300 &  76 & 36 & 50 & 65 & 27 & 39 & 53 & 19 & 30 \\
10 & 200 &  89 & 50 & 63 & 82 & 39 & 54 & 71 & 29 & 43 \\
10 & 300 &  82 & 44 & 57 & 74 & 34 & 48 & 63 & 25 & 38 \\
10 & 400 &  74 & 38 & 51 & 65 & 30 & 42 & 55 & 22 & 33 \\
\hline
\end{tabular}
\end{center}
\caption{ Efficiencies (in \%) for a cut on the $D_s$ momentum
at 2, 2.05 and 2.1 $GeV/c$ for four sets of values for the parameters
of the ACCMM model.}
\label{tab:eff_ds}
\end{table}
%%%%%%%%%%%%%%%%%%% Figure %%%%%%%%%%%%%%%%%%%%%%%%%% \setlength{\unitlength}{0.7mm}
\begin{figure*}[htb]
\vfill
\begin{picture}(280,90)(-25,-0)
%\begin{center}
%\FigureAng{vubdsmom.eps}{5.0cm}{10.0cm}{-90}
%\FigureAng{babardwg.eps}{9.0cm}{12.0cm}{90}
\mbox{\epsfig{file=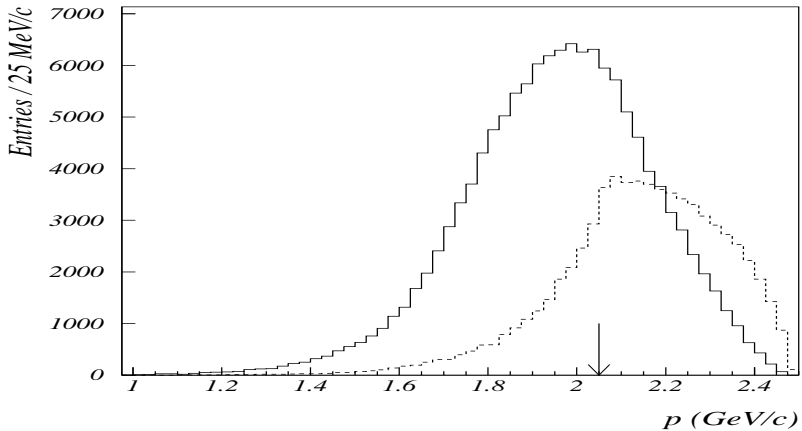,height=7.0cm,width=12cm}}
%\end{center}
\end{picture}
\vfill
\caption{Expected momentum spectrum for $D^+_s$ mesons produced from 
the reaction $\overline{b}\rightarrow D^+_s \overline{u}$ 
(i.e. upper vertex). The dashed line is for direct $D^+_s$ while the solid
line is for $D^+_s$ mesons coming from direct $D^{*+}_s$ decays.}
\label{fig:vubdsmom}
\end{figure*}
%%%%%%%%%%%%%%%%%%%% Fin Figure %%%%%%%%%%%%%

\section{Backgrounds with tagged events}
\hspace{\parindent}
Let us now discuss in more detail, the issues raised using this method.
The main questions are:
\begin{itemize}
\item Since we are not dealing with free quarks but B mesons, to which
extend does the factorization for the decay $B\rightarrow D_s^+ X$ hold ?
\item Are there other means to produce $D_s$ mesons~? 
\end{itemize}

The former question will not be addressed here beyond repeating 
that factorization seems to hold for the color allowed decays when 
confronting the data. Furthermore, we have computed \cite{7r} $O(\alpha_s)$ corrections to the
naive formula (\ref{eq:width}). \par

Other production sources of $D_s$ are shown in figure~\ref{fig:diagbkg}.
In the following, we discuss these various $D_s$ production mechanisms,
evaluate their rate and propose means to reject the ones involving 
$\overline{b} \rightarrow \overline{c}$ transitions or correct for the others. We
should distinguish between the background that concerns tagged or untagged events.
Let us begin here with tagged events. In $e^+e^- \to \Upsilon (4S) \to B
\overline{B}$, assume that the $\overline{B}$ is identified through its semileptonic
decay. Then, the right sign $D_s^+$ can be produced, besides the main mechanism of
Fig.~1, by mechanisms of Figs.~4(a)-(d). 
%%%%%%%%%%%%%%%%%%% Figure %%%%%%%%%%%%%%%%%%%%%%%%%% 
\setlength{\unitlength}{0.7mm} \begin{figure*}[htb] \vfill
\begin{picture}(280,140)(-35,-140) %\begin{center}
\FigureAng{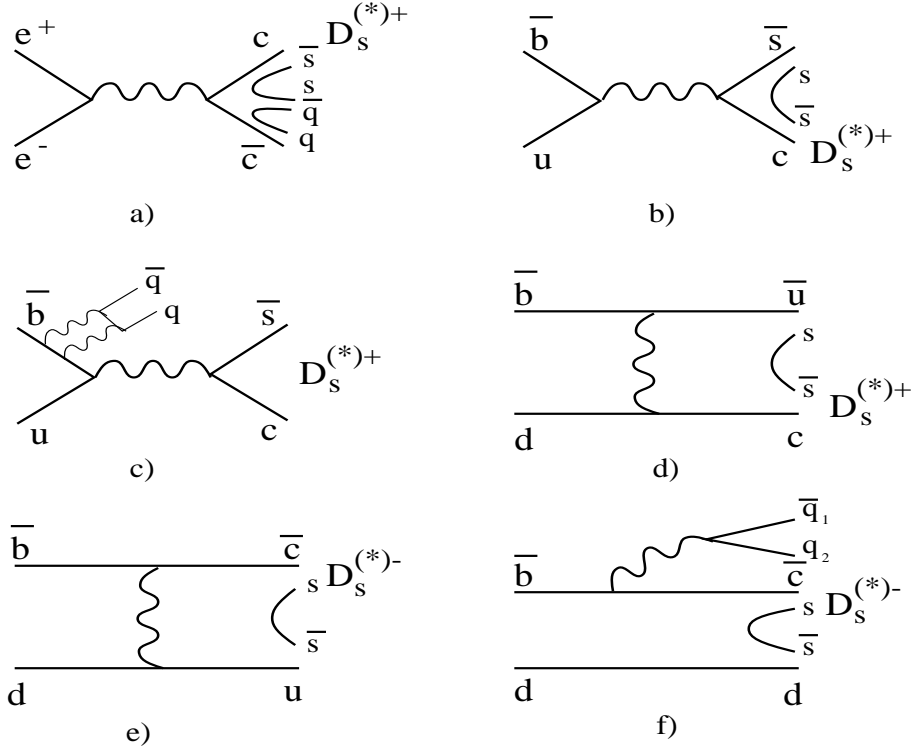}{10.0cm}{12.0cm}{-90}
%\FigureAng{babardwg.eps}{9.0cm}{12.0cm}{90}
%\mbox{\epsfig{file=/home/usr201/mnt/aleksan/btods_fig1.eps}} %\end{center}
\end{picture}
\vfill
\caption{Diagrams leading to the production of $D_s$ mesons. }
\label{fig:diagbkg}
\end{figure*}
%%%%%%%%%%%%%%%%%%%% Figure Unitarity Triangle %%%%%%%%%%%%%

The $c\overline{c}$ {\it continuum} background (Fig.~\ref{fig:diagbkg}a) 
has a large cross section,
$\sim 1.1 nb$. However, these events tend to have a jet-like structure and
therefore can be rejected to a large extend by topological cuts.
Furthermore, since a $D_s$ meson has to be produced, 
the creation of a $s\overline{s}$
pair is required, reducing the rate by about an order of magnitude. 
In addition, the momentum spectrum of the $D_s$ meson 
produced in the continuum has a mean value larger than $m_B/2$ reducing 
further this background by more that a factor 3.   
Finally, it is possible to substract the remaining background by taking data
just below the threshold for $B\overline{B}$ production. \par

The {\it annihilation} diagram in Fig.~\ref{fig:diagbkg}b is obtained 
from the calculation of the inclusive rate
$B^+ \rightarrow c\overline{s}$ using:
\be
J_{\mu q} = \overline{s}\gamma_\mu (1-\gamma_5) c 
\label{eq:jqa}
\ee
\be
\langle B^+ | A^\mu | 0\rangle = -i f_B p^\mu_B
\ee
\noi that gives \cite{10r}
\beqn
\Gamma^{(0)}(B^+\rightarrow c\overline{s}) & = &
\frac{N_c G_F^2}{8\pi} |V^*_{ub}V_{cs}|^2 f_{B}^2 m_B 
(m_c^2+m_s^2) 
\left( 1 - \frac{(m_c^2-m_s^2)^2}{m_B^2(m_c^2+m_s^2)} \right) \nonumber \\
& &\times 
\sqrt{1 - 2\frac{(m_c^2+m_s^2)^2}{m_B^2} + \frac{(m_c^2-m_s^2)^2}{m_B^4}} \ 
a^2_1
\eeqn
Neglecting the s quark mass, one gets
\be
\Gamma^{(0)} (B^+\rightarrow c\overline{s}) \simeq 
\frac{N_c G_F^2}{8\pi} |V^*_{ub}V_{cs}|^2 f_{B}^2 m_B m_c^2
\left(1 - \frac{m_c^2}{m_B^2} \right)^2 a^2_1
\ee
Taking into account that one needs to create a $s\overline{s}$ pair in order 
to obtain a $D^{(*)}_s$ meson, one can assume
\be
\Gamma^{(0)} (B^+\rightarrow D^{(*)+}_s X) \leq 
\frac{1}{3} \Gamma^{(0)} (B^+\rightarrow c\overline{s}) 
\ee

\noindent Since $m_B^2 f_B \cong m_D^2 f_D$ in the heavy quark limit the suppression
factor of this mechanism relative to the spectator quark model (\ref{eq:width}) will
be of the order or smaller than
  \be
{N_c \over 3} \left ( {m_c \over m_b} \right )^3 \sim 3 \times 10^{-2}
\ee
This branching fraction is small compared to the one deduced from 
Fig.~\ref{fig:diagsp}
and would represent a small correction. The contribution 
from the diagram in Fig.~\ref{fig:diagbkg}c 
requiring the coupling via 2 gluons is expected to be much smaller 
and can be neglected. \par

The {\it exchange} diagram shown in Fig.~\ref{fig:diagbkg}d is
evaluated in the
same way than the annihilation one using: 
\be
J_{\mu q} = \overline{q_2}\gamma_\mu (1-\gamma_5) q_1
\label{eq:jqe}
\ee
\be
\langle B^0 | A^\mu | 0\rangle = -i f_B p^\mu_B
\ee
and replacing $a_1$ with $a_2=c_2+c_1/N_c$ (color-suppressed process)
\be
\Gamma^{(0)} (B^o\rightarrow q_1\overline{q_2}) \simeq 
\frac{N_c G_F^2}{8\pi} |V^*_{q_2b}V_{q_1d}|^2 f_{B}^2 m_B m_c^2
\left(1 - \frac{m_c^2}{m_B^2} \right)^2 a_2^2
\label{eq:gcu}
\ee
where $q_1\overline{q_2}$ can either be $c\overline{u}$ or $u\overline{c}$.
One should keep in mind that in this case the factorization Ansatz is on
much weaker ground. 
Obviously, the case with $q_1=c$ and $\overline{q_2}=\overline{u}$ is 
suppressed since the CKM factors are $|V^*_{ub}V_{cd}|$. This means that this
mechanism in the case of tagging is Cabibbo suppressed and color suppressed relatively to
the main mechanism of Fig. 1. Comparing (\ref{eq:width}) to (\ref{eq:gcu}), the
reduction factor is of the order 

\be
\label{30e}
\tan^2 \theta_c \ {N_c \over 3} \left ( {m_c \over m_b} \right )^3 \left ( {a_2 \over a_1}
\right )^2 \simleq 10^{-4} \ee

\noindent and we can safely neglect this mechanism. The conclusion is that, if
there is tagging, the mechanisms that can compete with the interesting process of
Fig. 1 either can be discarded by kinematical cuts or are smaller by a factor of the
order $10^{-2}$. The method seems therefore safe if there is tagging. \par  

\section{Backgrounds with untagged events}
\hspace{\parindent} If in $e^+e^- \to B \overline{B}$ we assume no tagging, besides
the additional mechanisms of Figs.~4 a-d, we can have also the processes 4 e-f, that
lead to a wrong sign $D_s$. First, one must remark that the {\it continuum}
background can also lead to a wrong sign $D_s$ (the lower $D$ in the diagram 4a),
but we know that one can dispose off of these events by topological cuts. Also, the
{\it exchange} process 4e, that corresponds to replacing in 4d $\bar{u} \to \bar{c}$
and $c \to u$, can lead to a wrong sign $D_s$. Unlike the case with  $q_1=u$ and
$\overline{q_2}=\overline{c}$ this process is in principle enhanced because  the CKM factors are
$|V^*_{cb}V_{ud}|$.  Similarly a $s\overline{s}$ pair is required to get a
$D^{(*)+}_s$ meson and therefore with $|a_2| = 0.2$ one obtains a naive suppression
factor smaller than  
\be 
{1 \over 3} \left | {V_{cb} \over V_{ub}} \right |^2 N_c \left (
{m_c \over m_b} \right )^3 \left ( {a_2 \over a_1} \right )^2 \cong 0.20 \quad .
 \ee 
Although this source is only present for neutral B decays and is smaller than  the
spectator diagram in Fig~\ref{fig:diagsp}, the corresponding branching fraction could
be non negligible, and its possible suppression relies on dynamical assumptions that
are not very reliable. This branching fraction  and (\ref{30e}) as well may further be enhanced by the
emission of gluon from the initial light quark. In this case 
\cite{BSS,FM}, the most important changes relative to (\ref{eq:gcu})
are the absence of the $m_c^2/m^2_B$ dependence due to helicity 
and the presence of the factor $f_B^2/m^2_d$ instead of $f_B^2/m^2_B$ 
due to the gluon radiation from the initial light quark. 
Therefore, gluonic emission may enhance the rate of the exchange diagram
by one order of magnitude if one uses 
$m_d = 300$ Mev/c$^2$ since the d quark must be interpreted as 
a constituent quark in this process. However, as pointed out in \cite{20r}, the
presence of the infrared sensitive parameter $1/m^2_d$ makes problematic a rigorous
perturbative estimation of this contribution. Furthermore, in the full inclusive decay,
according to Heavy Quark Theory, this type of contributions should be suppressed by a
factor $1/m_b^3$ relative to the main spectator diagram. On the other hand, present limits
(for example $Br(B^o\rightarrow D_s^- K^+)< 2.4  \times 10^{-4}$ \cite{PDG}) tend to
disfavor a large enhancement. The same conclusion can be reached using D lifetime
measurements \cite{IB}. It is nevertheless important to find a way to either measure it or
to eliminate it. One possibility could be to observe some of these final states, for
example $D^{(*)-}_s K^{(*)+}$ and evaluate their contribution. 

The CLEO collaboration has measured the rate $\bar{b} \to D_s^-X$ \cite{cleo:dsl} due to
diagrams 4e-4f, although other sources exist (see next subsection). The total rate was found 
to be $(2.1\pm 1.0)$ \%. However the momentum spectrum of those $D_s$ is
expected to be rather soft with less than 0.5 \% 
of those having a momentum greater
than 2.0 GeV/c. This leads to an effective branching fraction
$Br(B\rightarrow D^{-}_s X\ [p_{D_s} > 2.0\ \mbox{GeV/c}]) 
< 1.5\times 10^{-4}$ at 90\% . \par

It is also possible to produce $D^{(*)}_s$ mesons of the wrong sign in {\it multibody
B decays} such  as the one shown in Fig.~4f. 
The decay rate of this type of modes 
is potentially large. However, one should note several important points.
\begin{itemize}
\item The production of $D^{**}$ with orbital excitation L=1 
would not lead to $D^{(*)}_s$ as this meson needs to be accompanied by a kaon
and the total mass $D^{(*)}_sK$ is larger than the $D^{**}$ mass. 
\item In the case of non resonant $D^{(*)}_sK$ production from the lower
vertex, the energy is shared between the final 3 or more
particles and therefore the momentum spectrum of the $D^{(*)}_s$ 
is softer and barely reaches
the range where $B\rightarrow D^{(*)+}_s X_u$ is expected. 
As discussed in the above subsection, CLEO measurements indicate 
that this type of decay should not be a problem.
\end{itemize} 

The CLEO measurement mentioned in the previous section shows that this
background should not be large.

\section{Feasibility study using the Babar detector}
\hspace{\parindent}
The feasibility of this new method for measuring $|V_{ub}|$ 
has been verified for the Babar detector at the SLAC B factory PEP-II.

We have used the full detector simulation and the
reconstruction program~\cite{BaBarReco} 
to generate 5000 events with the following decay of one B meson 
$ B \rightarrow D^{(*)} D^{(*)}_s$,
where $D_s$ decays to the $ \phi \pi$ final state, and 
$\phi \rightarrow K^+ K^-$. 
This gives a $D_s$ spectrum peaked at $1.7 ~ GeV/c$ in the center of mass
system, therefore only slightly below the expected signal of $D_s$
coming from $V_{ub}$ transitions.
Generic $B \bar B $ decays were used to measure 
the background level.

We have studied two crucial points for this analysis :
\begin{itemize}
\item the reconstruction efficiency for the $D_s$,
\item the momentum resolution.
\end{itemize}

The analysis to isolate the $D_s $ signal proceeds as following : 
$K^{\pm}$ 
 are identified using the combined information coming from the Silicon 
Vertex Tracker, the Drift Chamber and the DIRC detector and then selected
if their invariant mass is in the $ 1020 \pm 10 \ MeV/c^2$ interval. 
A third track,
assumed to be a pion, is then selected. A cut
on $\cos \psi$, $|cos \psi | > 0.4 $, where  $\psi$  is 
the angle between one Kaon and the $D_s$ momentum 
in the $\phi$ rest frame, is then applied.

%%%%%%%%%%%%%%%%%%% Figure %%%%%%%%%%%%%%%%%%%%%%%%%%
\setlength{\unitlength}{0.7mm}
\begin{figure*}[htb]
\vfill
\begin{picture}(280,160)(0,60)
%\begin{center}
%\FigureAng{vub_ds_rec.eps}{5.0cm}{10.0cm}{-90}
\mbox{\epsfig{file=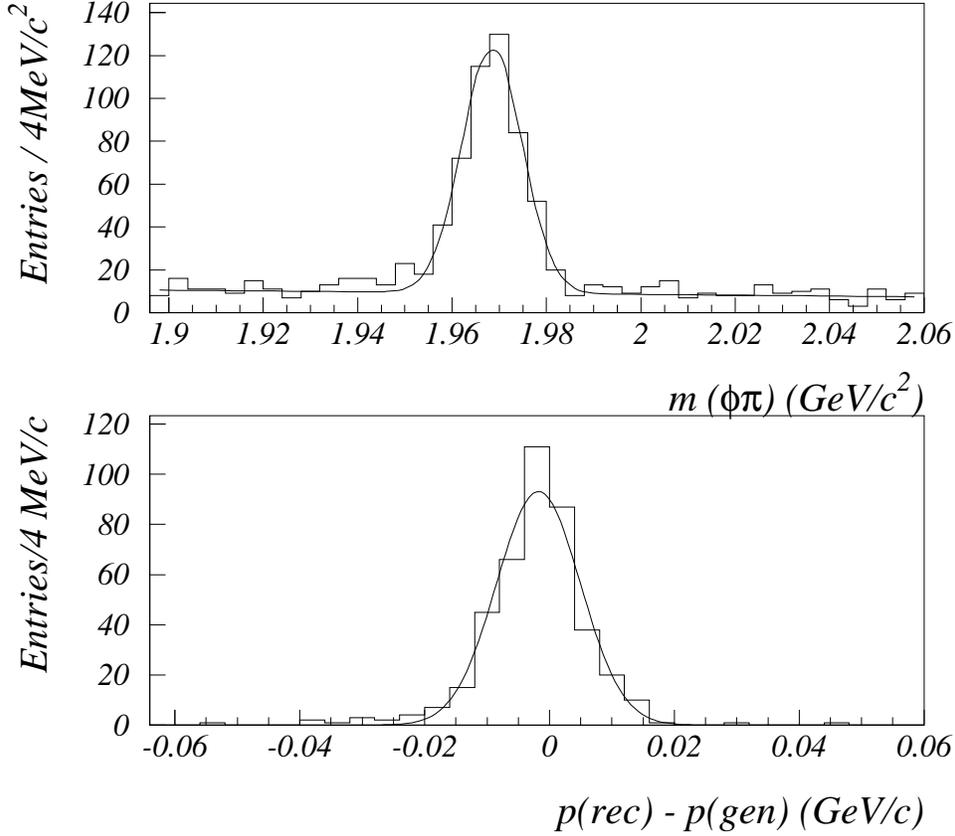,height=14.0cm}}
%\end{center}
\end{picture}
\vfill
\caption{Mass and momentum resolution for $D_s$ mesons using the full BaBar
detector simulation and reconstruction programs. }
\label{fig:DsReco}
\end{figure*}
%%%%%%%%%%%%%%%%%%%% Figure Unitarity Triangle %%%%%%%%%%%%%

The resolution on the $D_s$ mass is $6.4 \pm 0.3 ~MeV/c^2$ 
(Fig.~\ref{fig:DsReco}a).

The reconstruction efficiency is $ 39 \pm 2 $  \% and the momentum 
resolution is $6.7 \pm 0.3 ~ MeV/c$ (Fig.~\ref{fig:DsReco}b). 
The latter result insures that there will be
no leaking from the lower to higher momenta. This excellent resolution
is due to the fact that we implicitly reject 
mismeasured tracks : these will not give
a $D_s$ candidate with the right invariant mass.

Using these results, we can compute the number of reconstructed signal
events that we expect. We have for untagged events
$$ n_{rec} = 2 ~n_{B \bar B} Br (B \rightarrow D_s^{(*)} X_u )
Br(D_s^+ \rightarrow \phi \pi^+) 
Br(\phi \rightarrow K^+ K^-) \epsilon_{rec} = 143  $$
per $30 fb^{-1}$, the nominal integrated luminosity for one year of data
taking at Babar. 
We took $Br (B \rightarrow D_s^{(*)} X_u ) = 6.8~ 10^{-4} $ from 
Eq.~\ref{eq:BtoDsu},
$ Br(D_s^+ \rightarrow \phi \pi^+) = 3.5 \%$, and
$\epsilon_{rec} $ takes into account also the cut on $D_s$ momentum
at $2.05 ~ GeV/c$.
Therefore we can conclude that the number of reconstructed events
will be sufficient to measure $|V_{ub}|$ with a good statistical
precision. This number can be improved by reconstructing the $D_s$ meson  
in other modes.

As we have pointed out above, there are unwanted sources of $D_s$ beyond the
kinematical limit for $B \rightarrow D_s^+ X_c$, and it is suitable to be able to reject
them experimentally. As we have emphasized, one way to do this is to tag the flavor of
the recoil B meson in the event, for instance by considering its
semileptonic decay. The correlation between the sign of the lepton and
the sign of the $D_s$ meson is opposite for $B\rightarrow D_s^+ X_u$
and for the transitions due to the exchange
diagrams of figure 4-e (as well as for multibody
B decays like figure 4-f).

This method has already been used by other experiments
like CLEO and Argus to study the lepton spectrum in
the B semileptonic decays. A cut on the angle between
the lepton and the $D_s$ meson allows to reject the
pairs due to a $D_s$ and a lepton from the same B meson. The only major problem of
this method is the further reduction of the selected sample it implies, which
should be no larger than 5-10\% of the number of reconstructed
events estimated above. Therefore this is a possibility which is open but it would
probably require a big experimental effort to
reconstruct the largest possible fraction of $D_s$ mesons to be really
viable.

\section{Conclusion}
\hspace{\parindent}
In conclusion, we have shown that the process $B \to D_s^+  X_u$ can allow the
determination of the CKM matrix element $|V_{ub}|$ in $e^+e^-$ collisions at the
$\Upsilon (4S)$, as in the BaBar experiment. If there is tagging of one $B$ meson,
the prospects are very good since the backgrounds to the main spectator model
mechanism, whose spectrum would allow the determination of $|V_{ub}|$, are either
suppressed by a factor of the order of $10^{-2}$, or can be disposed off by
kinematical cuts. However, the number of events is drastically reduced by tagging.\par

 If tagging is not assumed, other mechanisms can give a large background, but the method
could still work if theoretical and experimental studies of these additional processes
leading to a wrong sign $D_s$ are performed in the future. It should be noted that these
wrong sign backgrounds (figures 4e and 4f) are Cabibbo enhanced but suppressed by color and
other dynamical effects (Section 5). In contrast, in semileptonic decays, even when the
hadronic background is studied \cite{1bisr}, misidentified direct $b \to c$ decays are Cabibbo
enhanced and difficult to exclude kinematically because of the neutrino. Admittedly, the
semileptonic method has the advantage of statistics.
 \par

 We are aware that our
study is a preliminary survey of the possibility of measuring $|V_{ub}|$ with a new method.
Work remains to be done. For the elementary processes $\bar{b} \to D_s^{(*)+}\bar{q}$
$(\bar{q} = \bar{u},\bar{c}$), one would need to compute the spectrum taking into account
the radiative corrections and comparison with the spectrum 
$B \to D_s^{\pm}X$
needs to be done as a check. Up to now, only the integrated corrected
rate has been computed \cite{7r}. On the other hand, theoretical or phenomenological work
needs to be done to further constrain the sources of background and of hadronic
uncertainties in the case of tagged and also, hopefully, although more difficult, for
untagged events. \par

We insist that the method proposed here, having very different systematic errors than the
semileptonic one, would provide an irreplaceable check of $|V_{ub}|$. 

\section*{Acknowledgements}
\hspace{\parindent} The authors acknowledge useful discussions with J.
Charles, and partial support from the EEC-TMR Program, contract N. CT 98-0169.

 \end{document}